\documentclass[prl,aps,graphicx,float]{revtex4}
\usepackage{epsfig}
\usepackage{bm}

\begin{document}

\preprint{}

\newcommand{\stef}[2]{$\blacktriangleright${\sc round #1:} {\em #2}$\blacktriangleleft$}

\title{Rectification in Luttinger liquids}

\author{D. E. Feldman$^{1}$, S. Scheidl$^{2}$, and V. M. Vinokur$^{3}$}

\affiliation{$^{1}$Department of Physics, Brown University, 
Providence, RI 02912
\\
$^{2}$Institut f\"ur Theoretische Physik, Universit\"at zu K\"oln,
Z\"ulpicher Strasse 77, D-50937 K\"oln, Germany
\\
$^{3}$Materials Science Division, Argonne National Laboratory, 9700
South Cass, Argonne, IL 60439}


\begin{abstract}
  
  We investigate the rectification of an ac bias in Luttinger liquids
  in the presence of an asymmetric potential (the ratchet effect). We
  show that strong repulsive electron interaction enhances the ratchet
  current in comparison with Fermi liquid systems, and the dc $I-V$ curve
  is strongly asymmetric in the low-voltage regime even for a weak
  asymmetric potential. At higher voltages the ratchet current
  exhibits an oscillatory voltage dependence.

\end{abstract}


\maketitle

Asymmetric conductors have asymmetric $I-V$ curves. This phenomenon is
known as the diode or ratchet effect and plays a major role in
electronics. Recently much interest has been attracted by transport
asymmetries in single-molecule devices and other mesoscopic systems
\cite{ratchets}.  The idea that asymmetric molecules can be used as
rectifiers is rather old \cite{smr}, however, it was implemented
experimentally \cite{sme} only recently. Another experimental
realization of a mesoscopic rectifier is an asymmetric electron
waveguide constructed within the inversion layer of a semiconductor
heterostructure \cite{heter}. The ratchet effect was observed in
carbon nanotubes \cite{nanotubes}, and strongly asymmetric $I-V$
curves were recently reported for the tunneling in the quantum Hall
edge states \cite{Hall}. These experimental advances have stimulated
much theoretical activity
\cite{mes-rat1,mes-rat2,mes-rat3,mes-rat4,mes-rat5} with the main
focus on the simplest Fermi-liquid systems \cite{Belinicher}.

Transport in one-channel quantum wires, where electrons form a
Luttinger liquid, differs significantly from the Fermi liquid case. In
particular, impurity effects are stronger in Luttinger liquids, and
even a weak impurity potential may render the linear conductance zero
at low temperatures \cite{KF}.  In this Letter we investigate the
ratchet effect in Luttinger liquids. We show that strong repulsive
electron interaction enhances the ratchet current, and the low-voltage
part of the $I-V$ curve is strongly asymmetric even in quantum wires
with weak asymmetric potentials.

We consider the ratchet effect in the presence of a weak asymmetric
potential $U(x)\ll E_F$, where $E_F$ is the band width. We calculate
the ratchet current $I_r(V)=[I(V)+I(-V)]/2$ for a one-channel quantum
wire with spin-polarized electrons. The ratchet current vanishes for
systems with symmetric $I-V$ curves. It can be measured as the dc
response to a low-frequency square voltage wave of amplitude $V$.
First, we consider voltages $V<V_0=\hbar v_F/(ea)$, where $v_F$ is the
Fermi velocity, $e$ the electron charge, and $a$ the size of the
region containing the asymmetric potential.  We find a weak ratchet
effect in the interval $eV_0> eV > \sqrt{U E_F}$ for both Fermi and
Luttinger liquids, $I_r\sim (e/h)U^2(eV)^{2g}/E_F^{2g+1}$, where $g=1$
for Fermi liquids and $g<1$ for Luttinger liquids with repulsive
interaction. However, at strong repulsive interaction (the Luttinger
liquid parameter $g\ll 1$) and sufficiently low voltages, the ratchet
current $I_r(V)$ {\it grows} as the voltage decreases until $I_r(V)$
becomes comparable with the total current $I(V)$ at
$eV=eV^*\sim(UE_F^{-g})^{1/(1-g)}$.  At $E_F\gg eV>eV_0$ the ratchet
current oscillates as a function of the voltage and can become
comparable with the total current $I(V)$ for any repulsive interaction
strength. We also briefly discuss the ratchet effect in the presence
of a strong asymmetric potential $U>E_F$. The complicated
ratchet-current behavior is caused by the energy dependence of the
effective impurity strength in Luttinger liquids \cite{KF}. This
introduces an additional energy scale $V^*$ absent in Fermi-liquid
systems.

One-channel quantum wires can be described by the Tomonaga-Luttinger
model with the Hamiltonian
\begin{equation} 
  \label{1} H=\int dx \bigg\{-\hbar v_F [ \psi_R^\dagger(x)
  i\partial_x\psi_R(x)- \psi_L^\dagger(x) i\partial_x\psi_L(x) ] 
  +  U(x) \rho(x)
  +\int dy
  K(x-y)  \rho(x) \rho(y) \bigg\}, 
\end{equation}
where $\psi_R^\dagger$ and $\psi_L^\dagger$ are the creation operators
for right- and left-moving electrons, $\psi^\dagger =
\psi^\dagger_R+\psi^\dagger_L$ gives the conventional electron
creation operator, $\rho=\psi^\dagger \psi$ is the electron density,
$U(x)$ is the asymmetric potential, and $K(x-y)$ the interaction
strength.  Our aim is to calculate the current $I$ as a function of
the applied voltage $V$.  We assume that the long-range Coulomb
interaction is screened by the gates so that $K(x-y)$ decreases
rapidly for large $(x-y)$.  Electric fields of external charges are
also assumed to be screened.  Thus, the applied voltage reveals itself
only as the difference of the electrochemical potentials $E_L$ and
$E_R$ of the particles injected from the left and right reservoirs.

We assume that one lead is connected to the ground so that its
electrochemical potential $E_R=E_F$ is fixed. The electrochemical
potential of the second lead $E_L=E_F+eV$ is controlled by the voltage
source. In such situation a symmetric potential $U(x)$ is sufficient
for rectification.  
For example, in a non-interacting system $I(V) \sim\int_{E_R}^{E_L}
[1-R(E)] dE$, where $R(E)$ is the reflection coefficient.  If the only
relevant scale for the energy dependence of the reflection probability
is the band width $\sim E_F$ then the ratchet current $
I_r\sim\int_0^{eV}dE [R(E_F-E)-R(E_F+E)]\approx-2\int_0^{eV}dE
R'(E_F)E\sim R(E_F)(eV)^2/E_F\sim U^2 (eV)^2/E_F^3$ for small $U$ and
$V$, and any coordinate dependence $U(x)$. 

A `not-trivial' ratchet effect can be observed when the injected
charge density is voltage-independent, $E_{L/R}=E_F\pm eV/2$. Symmetry
considerations require an asymmetric $U(x)$ for a non-vanishing
ratchet current in this case.  Also an electron interaction must be
present. Indeed, for free particles the reflection coefficient $R(E)$
is independent of the electron propagation direction \cite{LL} and
hence $I(V)=-I(-V)$.

The `non-trivial' ratchet effect is absent in the first two orders in
$U(x)$. Indeed, in the lowest two orders the ratchet current
$I_r^{(1,2)}=\int dx C(x) U(x)+ \int dx dy D(x,y) U(x) U(y)$.
$I_r^{(1,2)}$ must be zero for any symmetric potential.  Substituting
$U(x)=U\delta(x-x_0)$ we find that $C(x_0),D(x_0,x_0)=0$.  Substituting
$U(x)=U\delta(x-x_1)+U\delta(x-x_2)$ we see that
$D(x_1,x_2)+D(x_2,x_1)=0$. Hence, $I_r^{(1,2)}=0$ for any $U(x)$.



We first consider the `non-trivial' ratchet effect and then check what
changes after the voltage dependence of the injected charge density is
taken into account.  Let us begin with a qualitative explanation
before we make a rigorous calculation. The origin of the ratchet
current can be understood from a simplified Hartree-Fock picture. In
this approximation, electrons are backscattered off a combined potential
$\tilde U(x)=U(x)+W(x)$, where $W(x)$ is a self-consistent
electrostatic potential created by the average local charge density.
To obtain $W(x)$ we use the following approximation in the last term of Eq. (\ref{1}):
$\rho(x)\rho(y)\approx (\psi_R^+(x)\psi_R(x)+\psi_L^+(x)\psi_L(x))(\psi_R^+(y)\psi_R(y)+\psi_L^+(y)\psi_L(y))+
[\langle\rho(x)\rangle\psi_R^+(y)\psi_L(y)+ \langle\rho(y)\rangle\psi_R^+(x)\psi_L(x) + h.c.]+{\rm const}$.
Thus, the relation between $W$ and $\rho$ is linear.
The combined potential $\tilde U(x)$ is different for the opposite
voltage signs. 



\begin{figure}
  \epsfig{file=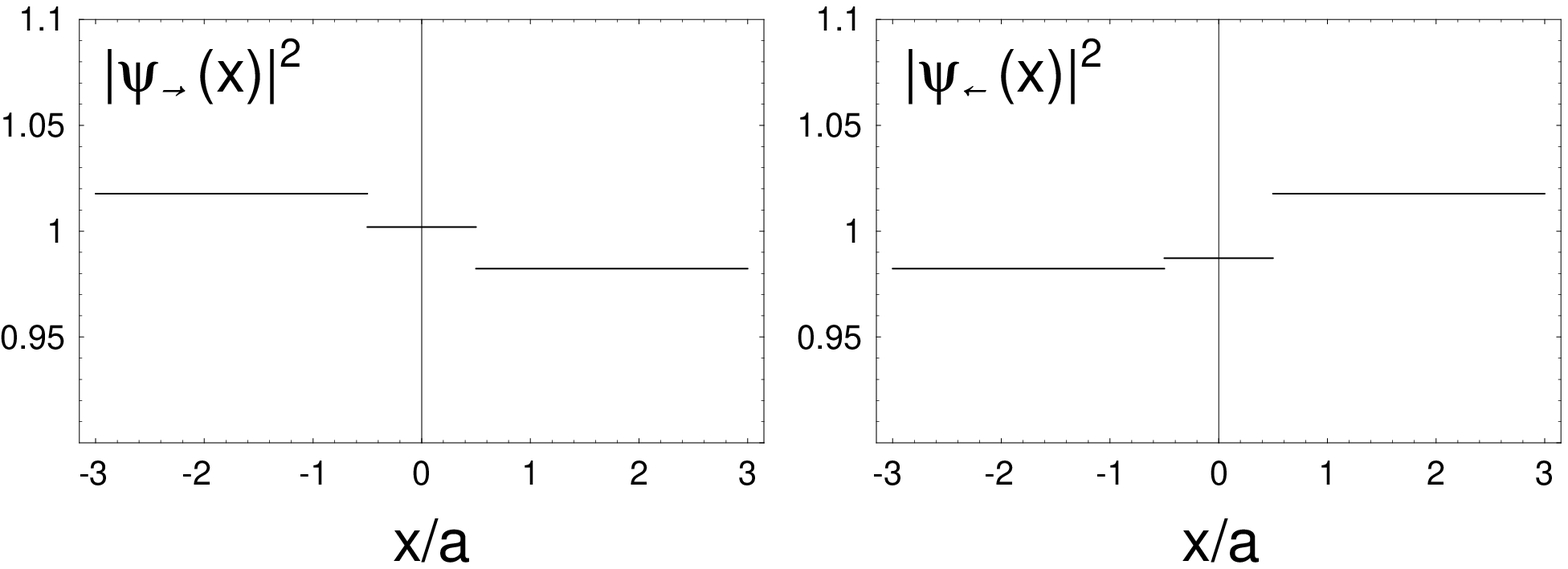, width=\linewidth} 
  \caption{%
    Density profiles averaged over the period of Friedel oscillations
    for a potential with $U_1<U_2$. The averaged densities show drops
    at the impurity positions.}
\label{fig}
\end{figure}

In the model (\ref{1}) the electron interaction is short-ranged due to
the screening gates, and hence, the relation between the potential
$W(x)$ and the electron density $\rho(x)$ is local, $W(x)\sim\rho(x)$. 
The simplest choice of
$U(x)$ is a two-impurity asymmetric potential
$U(x)=U_1\delta(x+a/2)+U_2\delta(x-a/2).$ The charge density profile
\cite{footnote1} in the presence of a two-impurity potential and the
voltage drop $V$ is sketched in Fig. \ref{fig}.
Depending on the voltage sign, the charge density decreases or grows
as a function of the coordinate $x$. So does the electrostatic
potential $W(x)$. Hence, $\tilde U(x)$ is different for the opposite
voltage signs. 
The density is essentially independent of the coordinate
between the impurities \cite{footnote1}, as well as on the left and on
the right of the impurities, since no backscattering occurs in those
regions.  The charge density and the electrostatic potential drop at
the positions of the impurities. The magnitude of a drop is
proportional to the electric charge backscattered off the impurity.
Indeed, if the incident charge densities of the electrons approaching
the impurity from the left and from the right are $\rho_L^\rightarrow$
and $\rho_R^\leftarrow$, and the backscattered charge densities are
$\rho_L^\leftarrow$ and $\rho_R^\rightarrow$ then the density drop
across the impurity $\Delta\rho = (\rho_L^\rightarrow +
\rho_L^\leftarrow + \rho_R^\leftarrow - \rho_R^\rightarrow) -
(\rho_R^\leftarrow + \rho_R^\rightarrow + \rho_L^\rightarrow -
\rho_L^\leftarrow)= 2(\rho_L^\leftarrow - \rho_R^\rightarrow) \sim
I_{\rm bs}$, where $I_{\rm bs}$ is the current backscattered off the
impurity. 
Thus $W(x)=\tilde U - U \sim I_{\rm bs}$. From Ref.  \cite{KF} we know
that for a weak potential $U$
\begin{eqnarray} 
  \label{3}I_{\rm bs}\sim |U_{2k_F}|^2|V|^{2g-1}{\rm
    sign}V/E_F^{2g},
\end{eqnarray} 
where $U_{2k_F}\sim k_F \int dx \exp(2ik_Fx) U(x)$, $k_F$ is
proportional to the mean electron density, and the dimensionless
constant $g$ characterizes the interaction strength, $g=1$ for
non-interacting electrons (in which case $W(x)=0$).
%

Now we can substitute the
renormalized potential $\tilde U=U+W$ for $U$ in Eq. (\ref{3}). The
Fourier component $W_{2k_F}$ is different for the opposite voltage
signs.  Hence, we obtain the asymmetric part of the $I-V$
characteristics
$I_{\rm r}\sim eU^3|eV|^{4g-2}/(h E_F^{4g})$.
The ratchet effect is strongest for $g\rightarrow 0$ when the ratchet
current grows as the voltage decreases.

The above Hartree-Fock argument provides a qualitatively correct
picture at small $g$ but underestimates fluctuations in Luttinger
liquids. As shown below, the ratchet current growth at small voltages
differs from our estimate: $I_r\sim U^3 |V|^{6g-2}$, $E_F\gg V>V^*\sim
(UE_F^{-g})^{1/(1-g)}$, $g\ll 1$.
We will see that the growth terminates at $V=V^*$. At such voltage
$I_r(V^*)/I(V^*)\sim [(V^*)^{3g+1}/E_F^{3g}]/[V^*]\sim
(V^*/E_F)^{3g}\sim 1$ as $g\ll 1$. Fluctuations are less important in
many-channel systems and the Hartree-Fock picture gives exact results
for some two-channel systems and for Fermi liquids \cite{unp}.

We use the bosonization technique \cite{bos} to calculate the ratchet
current.  After an appropriate rescaling of the time variable, the
system can be described by the action \cite{KF}
\begin{eqnarray}
  \label{5}
  S&=&\int
  dtdx\bigg\{\frac{1}{8\pi}[(\partial_t\Phi)^2-(\partial_x\Phi)^2]
    -\delta(x)\sum_{n \geq1}
    2 \tilde U_{2nk_F} \cos(n\sqrt{g}\Phi+\alpha_n)\bigg\}, 
\end{eqnarray} 
where the bosonic field $\Phi$ is related to the charge density as
$\rho=e(\sqrt{g}\partial_x\Phi+2k_F)/(2\pi)$, and $\tilde
U_{2nk_F}\exp(i\alpha_n)$ are of the order of the Fourier components
of the asymmetric potential, $k_F\int\exp(2ink_Fx) \tilde U(x)dx$. We
assume that the charge density $\sim k_F$ is independent of the
voltage.
The operator $\cos(n\sqrt{g}\Phi+\alpha_n)$ describes scattering
events involving $n$ electrons. We assume that $\alpha_1=0$.  Indeed,
we can always set $\alpha_1=0$ by a constant shift of the bosonic
field $\Phi$. For a general asymmetric potential, $\alpha_n$ with
$n>1$ remain non-zero after this shift.  On the other hand, for a
symmetric potential $U(x)=U(-x)$ all $\alpha_n=0$.  In
most problems it is sufficient to keep only the $n=1$ term. The $n=2$
contribution is relevant in the theory of resonant transmission in
Luttinger liquids \cite{KF}. This term is also important for the
ratchet effect.

We use the standard model \cite{leads,FG,TSDG} for Fermi-liquid leads
adiabatically connected to the wire. We assume that the action
(\ref{5}) is applicable for $|x|<L$ only. At large $|x|$ the
interaction strength $K(x-y)$, Eq. (\ref{1}), is zero.  This model can
be interpreted as a quantum wire with electron interaction completely
screened by the gates near its ends. Electric fields of external
charges are assumed to be screened in all parts of the wire.  A simple
modification of this model describes electrically neutral leads
\cite{TSDG}.  All results coincide for our set-up and the model
\cite{TSDG}.

The current injected from the non-interacting 1D regions is given by
the Landauer formula $I_0=e^2V/h$ \cite{leads}. Indeed,
left-/right-movers entering the non-interacting region from the
central part of the wire cannot affect the current of
right-/left-movers in the non-interacting region. Hence, the current
of right-/left-moving particles in the left/right non-interacting
region is determined by the chemical potential of the left/right
reservoir.  The total current is the sum \cite{FG,TSDG} of the
injected current and the current backscattered off the asymmetric
potential: $I=I_0+I_{\rm bs}$.  Only $I_{\rm bs}$ contributes to the ratchet
effect. To find the backscattered current we employ the Keldysh
formalism \cite{Keldysh}. We assume that at $t=-\infty$ there is no
backscattering in the Hamiltonian, and then the backscattering is
gradually turned on. Thus, at the initial moment of time the numbers
$N_L$ and $N_R$ of left- and right-moving electrons conserve
separately. Hence, at $t=-\infty$ the system can be described by a
partition function with two chemical potentials $\mu_R=E_F$ and
$\mu_L=E_F+eV$ conjugated with the particle numbers $N_R$ and $N_L$.
This initial state determines bare Keldysh Green functions.

We will consider only zero temperature.  It is convenient to switch
\cite{FG} to the interaction representation $H\rightarrow H-\mu_R N_R
- \mu_L N_L$. This transformation induces time dependence in the
electron creation and annihilation operators. As a result, $\sum_n 2
\tilde U_{2nk_F} \cos(n\sqrt{g}\Phi+\alpha_n)$ in the action should be
modified as $\sum_n 2 \tilde U_{2nk_F} \cos(n\sqrt{g}\Phi+\alpha_n+nA(t))$,
where $A(t)=eVt/\hbar$ \cite{KF,FG,TSDG}. The backscattered current
operator equals \cite{KF,TSDG}
\begin{equation}
  \label{6}
  I_{\rm bs}(t)=dN_L/dt=i[H,N_L]/\hbar=-\delta S/\delta A(t),
\end{equation}
where we omit dimensional factors such as $e$, $\hbar$, and $v_F$ for brevity.
We need to calculate
\begin{eqnarray}
  \label{7}
  \langle \hat I_{\rm bs}(t=0)\rangle=
  \langle 0|S(-\infty;0) \hat I_{\rm bs} (0)
  S(0;-\infty) |0\rangle, 
\end{eqnarray} 
where $|0\rangle$ denotes the initial
state and $S$ is the evolution operator. In the weak impurity case
this can be done with the perturbation theory in $\tilde U_{2nk_F}$ using the
bare Green function \cite{bos} $\langle 0|\Phi(t_1,x_1=0) \Phi(t_2,
x_2=0)|0\rangle=-2\ln(\delta+i[t_1-t_2])$, where $\delta$ is an
infinitesimal positive constant.

If all $\alpha_k=0$ then the ratchet current is zero. Indeed, at
$\alpha_k=0$ the action (\ref{5}) is invariant under the
transformation $\Phi\rightarrow-\Phi$, $V\rightarrow -V$ while the
current operator (\ref{6}) changes its sign.
As discussed above, for an asymmetric potential we expect $\alpha_2\ne
0$. Then a ratchet current $I_r$ emerges in the order $\tilde
U_{2k_F}^2\tilde U_{4k_F}$. Before the calculation of $I_r$ let us
determine its voltage dependence with a heuristic argument similar to
Ref. \cite{KF}. As one changes the energy scale $E$, the
backscattering amplitudes $\tilde U_{2nk_F}$ in the action (\ref{5})
scale as $\tilde U_{2nk_F}(E) \sim \tilde U_{2nk_F} E^{n^2 g -1}$
\cite{KF}. This renormalization stops at the energy scale $V$.
Assuming that a scattering matrix approach could be applied for an
estimation of the current, we write $I_{\rm bs}(V)\sim V R_{\rm
  eff}(V)$, where $R_{\rm eff}(E)= \sum {\rm const} \tilde
U_{2nk_F}^2(E) + \sum{\rm const}\tilde U_{2nk_F}(E) \tilde
U_{2mk_F}(E) \tilde U_{2lk_F}(E) +\dots$ is an effective reflection
coefficient. Quadratic terms do not contribute to the ratchet current.
The leading contribution emerges in the order $\tilde U_{2k_F}^2\tilde
U_{4k_F}$. One gets $I_r\sim V \tilde U_{2k_F}^2(V)\tilde
U_{4k_F}(V)\sim V^{6g-2}$.
Below we obtain the same result rigorously from Eqs. (\ref{6},\ref{7}).

Expanding Eq. (\ref{7}) to the order $\tilde U_{2k_F}^2\tilde U_{4k_F}$ gives
\begin{widetext}
\begin{eqnarray}
  \label{8}
  I_r&=&2\sin\alpha_2 \tilde U_{2k_F}^2 \tilde U_{4k_F} \bigg\{ \int_{-\infty}^0
  dt_1\int_{-\infty}^{t_1}dt_2 \cos(Vt_1-2Vt_2) P(t_1, t_2, t_2-t_1)
  \nonumber\\&&
  +\int_{-\infty}^0 dt_1 \int_{t_1}^0 dt_2 \cos(Vt_1-2 Vt_2) P(t_1, t_2,
  t_1-t_2) 
  - \int_{-\infty}^0 dt_1\int_{-\infty}^0 dt_2
  \cos(Vt_1-2Vt_2)P(-t_1,t_2-t_1, t_2)
  \nonumber\\&&
  - 2\int_{-\infty}^0 dt_1\int_{-\infty}^{t_1} dt_2 \cos(Vt_1+Vt_2)
  P(t_2-t_1, t_1, t_2) 
  +\int_{-\infty}^0 dt_1\int_{-\infty}^0 dt_2 \cos(
  Vt_1+Vt_2) P(t_2-t_1,-t_1, t_2) \bigg\} + c.c., 
\end{eqnarray} 
where $P(t,s,q)=(\delta-it)^{2g}(\delta-is)^{-4g}(\delta-iq)^{-4g}$.
Dimensional analysis shows that $I_r\sim V^{6g-2}$ in agreement with
our previous estimate. It is convenient to change variables in the
integrals with $\cos( V t_1- 2 V t_2)$ as $\tau_1=t_2-t_1$,
$\tau_2=t_2$.  Then after tedious but straightforward manipulations
Eq. (\ref{8}) can be represented as
\begin{eqnarray}
  \label{9}
  I_r=2\sin\alpha_2 \tilde U_{2k_F}^2 \tilde U_{4k_F} 
  \left \{ \int_{-\infty}^{\infty}
    dt d\tau \cos [V(\tau - t)] P(-(t+\tau),-\tau,-t) -
    \int_{-\infty}^{\infty} dt d\tau \cos [V(\tau+t)]
    P(\tau-t,\tau,t)\right\}.  
\end{eqnarray} 
\end{widetext}
The first integral in (\ref{9}) is
zero as seen from the location of the branching points of the function
$P$. The second integral yields
\begin{eqnarray}
  \label{10}
  I_r=-\sin\alpha_2\tilde U_{2k_F}^2 \tilde U_{4k_F} \cos(\pi
  g)\frac{2^{2+2g}\pi^{3/2}\Gamma (g+1/2)} {\Gamma (4g) \Gamma (3g)}
  |V|^{6g-2}.  
\end{eqnarray} 

This expression becomes $0$ at $g=1/2$. We also get a zero ratchet
current for non-interacting electrons, $g=1$, because the Hamiltonian
(\ref{1}) is quadratic in Fermi-operators in the non-interacting case
and hence no operators which backscatter more than one electron can appear, 
$\tilde U_{4k_F}=0$.

At small $g$ the ratchet current (\ref{10}) is proportional to a
negative power of the voltage.  This means an unusual behavior: the dc
response to an ac voltage grows as the ac voltage decreases.

So far we ignored the voltage dependence of the injected charge
density.  At $g\ll 1$, Eq. (\ref{10}) gives the main contribution to
the ratchet current only for $eV<\sqrt{U E_F}$. For $g$ close to $1$
the result (\ref{10}) is always exceeded by another contribution.
This contribution emerges in the second order in $U$ and is related to
the voltage dependence of the injected charge density. The density is
proportional to $k_F$ which enters the expression for $U_{2k_F}$ in
Eq. (\ref{3}). At small $V\ll E_F$ the correction \cite{footnote2} to
$U_{2k_F}$ is a linear function of $V$. The substitution of this
correction into Eq. (\ref{3}) gives an additional ratchet current
\begin{eqnarray}
  \label{11}
  I_r^{\rm (density)}\sim \frac{e U_{2k_F}^2 (eV)^{2g}}{hE_F^{2g+1}}.
\end{eqnarray}

For $g>1/3$ and $V>V^*$ the contribution (\ref{11}) always exceeds (\ref{10}).  At
$g<1/3$ the current (\ref{11}) is greater than (\ref{10}) above a
threshold voltage that depends on $U$ and $g$.  As we already
discussed, $I_r$ (\ref{10}) is comparable with the total current
$I(V)\sim e^2V/h$ at small $g$ near the border of the perturbatively
accessible region $UV^{g-1}/E_F^g<1$. 
On the other hand, Eq.
(\ref{11}) provides only a small correction to the total current
for any $g$. Still a repulsive interaction of any strength enhances
the ratchet effect as seen from the comparison of the current
(\ref{11}) for $g<1$ and for the non-interacting case $g=1$.

What happens beyond the perturbative region when $V<V^*\sim U
^{1/(1-g)}$? As the energy scale decreases the effective impurity
strength grows. Hence, we need to consider a strong $U>E_F$ limit. In
this limit we have a weak tunneling between the left and right halves
of the wire.  The current $I(V)\sim t^2 V^{(2/g)-1}/E_F^{2/g}$, where
$t$ is the tunneling amplitude \cite{KF}. Inserting the voltage
dependence of the tunneling amplitude in the expression above we
estimate $I_r(V)\sim V^{2/g}$.

A single impurity model (\ref{5}) can be used only 
when the potential $U(x)$ is confined in a small
space region of size $a<a_V\sim\hbar v_F/(eV)$.  If the potential
changes slowly at the scales $x>a_V\gg 1/k_F$ it cannot backscatter
electrons since backscattering involves high momentum transfers,
$\Delta k \ge k_F$. Interesting interference effects are possible for
a two-impurity potential $U_1\delta(x)+U_2\delta(x-a)$ and other
$U(x)$ which significantly change at the scale $1/k_F$ but are
non-zero in a region of size $a\sim a_V$. In the two-impurity case the
current oscillates as a function of the voltage bias
\cite{interferometry}. For $U_1,U_2\ll E_F$, $I-e^2V/h\sim
[U_1^2+U_2^2+2U_1U_2 \cos(2k_Fa)H(geVa/[\hbar v_F])]|V|^{2g-1}{\rm
  sign} V$, where
$H(x)=\sqrt{\pi}\Gamma(2g)J_{g-1/2}(x)/[\Gamma(g)(2x)^{g-1/2}]$ and
$J_{g-1/2}(x)$ is the Bessel function of the first kind
\cite{interferometry}.  The main contribution to the ratchet current
at $a\sim a_V$ comes from the shift of $k_F$ due to the change of the
electrochemical potential of the left reservoir by $eV$.
{From} the minimum of the quadratic part of the bosonized Hamiltonian
one finds the charge density shift \cite{G}.  This gives $k_F=k_F^{(0)}+g^2
eV/(2\hbar v_F)$. After the substitution to the expression for the
total current $I$ we find
\begin{eqnarray}
  \label{12}
  I_r(V) &\sim& U_1U_2\sin(2k_F^{(0)}a)|V|^{2g-1} 
  \sin(g^2e|V|a/[\hbar v_F])H(geVa/[\hbar v_F])
\end{eqnarray}
Thus, $I_r(V)$ oscillates. Notice that for $V\sim V^*\ll E_F$, $a\sim a_{V^*}$ the
ratchet current (\ref{12}) is of the order of the total current $\sim
e^2V/h$.

In conclusion, we have found the ratchet current for strong and weak
asymmetric potentials. It exhibits a set of universal power
dependencies on the voltage and can grow as the voltage decreases.
This work was supported by the US DOE Office of Science under contract
No.  W31-109-ENG-38.

\newpage

\begin{thebibliography} {150}
\bibitem{footnote1} We do not take Friedel oscillations into account
  since they do not change our estimate qualitatively \cite{unp}.
\bibitem{footnote2} The applied voltage renormalizes other terms in
  the action too (for example, the Luttinger parameter $g$). Power
  counting shows that some of them could also give rise to ratchet
  current contributions of the order of $U^2V^{2g}\ln^{s} V$, $s={\rm
    const}$.
\end{thebibliography}
\end{document}